\documentclass[12pt,preprint]{aastex}

\def\fermi{{\it Fermi}}

\usepackage{lscape}

\usepackage{epstopdf} % to include .eps graphics files with pdfLaTeX
\usepackage{graphicx}
\usepackage{epsfig}
\usepackage{natbib}
\usepackage{amsmath,amssymb}  % Better maths support & more symbols

\usepackage{ifpdf}

\ifpdf
\usepackage[pdftex,usenames,dvips]{color}
\else
\usepackage[usenames,dvips]{color}
\fi

\begin{document}

\title{Orbital Phase Dependent $\gamma$-ray emissions from the Black Widow Pulsar}
\author{E.~M.~H.~Wu\altaffilmark{1},
%\email{wmheric@gmail.com}
J.~Takata\altaffilmark{1},
%\email{takata@hkucc.hku.hk}
K.~S.~Cheng\altaffilmark{1},
%\email{hrspksc@hkucc.hku.hk}
R.~H.~H.~Huang\altaffilmark{2},
C.~Y.~Hui\altaffilmark{3},
A.~K.~H.~Kong\altaffilmark{2},
P.~H.~T.~Tam\altaffilmark{2},
and
J.~H.~K.~Wu\altaffilmark{2}
}
\altaffiltext{1}{Department of Physics, University of Hong Kong, Pokfulam Road, Hong Kong}
\altaffiltext{2}{Institute of Astronomy and Department of Physics, National Tsing Hua University, Hsinchu, Taiwan}
\altaffiltext{3}{Department of Astronomy and Space Science, Chungnam National University, Daejeon, Republic of Korea
}

\email{wmheric@gmail.com, takata@hku.hk, hrspksc@hkucc.hku.hk}

\begin{abstract}
We report on evidence for orbital phase-dependence of the $\gamma$-ray emission from PSR~B1957+20 black widow system by using the data of the \fermi\ Large Area Telescope. We divide an orbital cycle into two regions: a region containing the inferior conjunction, and the other region containing rest of the orbital cycle. We show that the observed spectra for the different orbital regions are fitted by different functional forms. The spectrum of the orbital region containing inferior conjunction can be described by a power-law with an exponential cutoff (PLE) model, which gives the best-fit model for the orbital phase that does not contain the inferior conjunction, plus an extra component above $\sim 2.7$~GeV. The emission above $3$~GeV in this region is detected with a $\sim7\sigma$ confidence level. The $\gamma$-ray data above $\sim 2.7~{\rm GeV}$ are observed to be modulated at the orbital period at the $\sim 2.3 \sigma$ level. We anticipate that the PLE component dominating below $\sim 2.7~{\rm GeV}$ originates from the pulsar magnetosphere. We also show that the inverse-Compton scattering of the thermal radiation of the companion star off  a ``cold'' ultra-relativistic pulsar wind can explain the extra component above $\sim$2.7~GeV. The black widow pulsar PSR~B1957+20 may be the member of a new class of object, in the sense that the system is showing $\gamma$-ray emission with both magnetospheric and pulsar wind origins.
\end{abstract}


\section{Introduction}
The $Fermi$ Large Area Telescope ($Fermi$) has measured the pulsed $\gamma$-ray emission from millisecond pulsars, which is magnetospheric in origin \citep{abd09a,abd09b,gui12a}. Furthermore, more millisecond pulsars (MSPs) have been discovered as $Fermi$ unidentified sources in radio bands (e.g., \citealt{cog11,ran11,gui12a,ker12,ray12}), suggesting MSPs are one of the major Galactic sources in the high-energy $\gamma$-ray sky. The pulsed $\gamma$-ray emissions from energetic MSP, PSR B1957+20, was reported by \citet{gui12b}. PSR B1957+20 has been known as the first ``black widow'' binary system, in which the MSP is destroying the low mass companion star ($<0.1 M_{\odot}$) \citep{fst88}.So far, at least 10 black widow systems have been discovered in the $Fermi$ unidentified sources \citep{rob11,ray12}.

\citet{sta03} detected the unresolved X-ray emission around PSR~B1957+20, and \citet{hua12} found the modulation of the unresolved X-ray emissions with the orbital  period of $P_o\sim 9.2$~hrs, suggesting the origin of an intra-binary shock \citep{har90,aro93,che06}. In the intra-binary shock scenario, the matter of the wind from the extremely low mass companion is injected by the irradiation of the strong wind/radiation from the MSP, and collides with the pulsar wind at a position not far from the companion \citep{che89}. The pulsed radio emissions from PSR~B1957+20 shows the eclipse lasting for about 10\% of the orbital phase \citep{fst88}. The corresponding size of the eclipse is on order of $\sim 10^{11}$~cm, which is larger than the Roche lobe size $\sim 2\times 10^{10}$~cm, which implies that stellar material is expelled from the companion by the pulsar wind \citep{rud89}. The irradiation of the pulsar wind/radiation on the companion star of PSR~B1957+20 system has also been expected by the observations of the orbital modulation of the optical emissions from the companion \citep{cpr95,rey07}. In addition to the PSR~B1957+20 system, the orbital modulation of the optical emission associated with the irradiation of the MSP has been confirmed for several binary systems, for example, the J1023+0038 system \citep{tho05}, the black widow system candidates 2FGL J2339.6-0532 \citep{rom11,kon12a} and 2FGL J1311.7-3429 \citep{kat12,rom12}, and the accreting MSPs in quiescent state, e.g. SAX J1808.4-3658 \citep{bur03,del08}.

The high-energy emission associated with an intra-binary shock between a pulsar and a companion star has been detected from the so-called $\gamma$-ray binary PSR~B1259-63/LS~2883 system, which is composed of a canonical pulsar with a period of $P\sim 48$~ms and a high-mass Be star \citep{joh92,aha09,uch09}. In the $\gamma$-ray binary, it has been proposed that the shock stands at the interface between the pulsar wind and the Be wind/disk, and un-pulsed radio to TeV radiations are produced via the synchrotron and inverse-Compton processes of the accelerated particles at the shock \citep{tav97,dub06,tak09,kon11,kon12b,tak12}. The flare-like GeV emissions have been detected by the $Fermi$ observation during the 2010/2011 periastron passage \citep{abd11,tam11}. Different models have been proposed for explaining the flare-like GeV emission from the PSR B1259-63/LS 2883 system: the Doppler boosting model \citep{kon12b} assumes that the flow in the bow-shock tail is relativistic and the synchrotron emission is Doppler-boosted, while the pulsar wind model \citep{kha12} assumes that the GeV emissions are produced by the inverse-Compton process of the cold-relativistic pulsar wind.

The $\gamma$-ray emissions associated with the pulsar wind in the black widow system PSR B1957+20 have not been reported yet. However, we note that for a typical $\gamma$-ray binary system, the distance to the shock from the pulsar is order of $R_{\rm s}\sim 0.1-1$~AU, corresponding to $R_{\rm s}/R_{\rm lc}\sim 10^{4}-10^{5}$, where $R_{\rm lc}=2.4\times 10^{8}(P/0.05~\rm s)~$cm is the light cylinder radius and $P$ is the spin period of the pulsar. For the black widow system, the shock stands at the distance $R_{\rm s}\sim 10^{10}-10^{11}$~cm, corresponding to $R_{\rm s}/R_{\rm lc}\sim 10^{3}-10^{4}$, which is roughly one order of magnitude smaller than that of the $\gamma$-ray binaries. Therefore, the black widow system will provide us a unique laboratory to probe the physics of the pulsar wind, and may be candidate of $\gamma$-ray emissions from the pulsar wind. In this paper, we report searching for orbital modulation of GeV $\gamma$-rays from PSR~B1957+20 system in the $Fermi$ data to obtain an evidence of the $\gamma$-ray emissions from the intra-binary space.

\section{Data analysis and results}
\label{data}
In this analysis, \fermi-LAT data were taken between 2008 August 4 and 2011 December 5. We restricted the events in the ``Source'' class (i.e., event class 2) under the ``P7SOURCE\_V6'' instrumental response functions (IRFs). In addition, we excluded the events with zenith angles greater than $100\degr$ to reduce the contamination by Earth albedo gamma-rays. The data were analyzed using the \fermi\ Science Tools v9r23p1, available from the \fermi\ Science Support Center \footnote{http://fermi.gsfc.nasa.gov/ssc/data/analysis/software/}. Events were selected within a circular region of interest (ROI) centered at the position of: R.A.~$=19^{\mathrm h} 59^{\mathrm m} 36\fs77$, decl.~$=20\degr 48\arcmin 15\farcs1$ \citep{aft94}. In order to reduce systematic uncertainties due to the surrounding complex galactic region and to achieve better background modelling, photon energies were restricted to be above $200$ MeV and a size of radius $5\degr$ of the ROI was adopted throughout the analysis. As an attempt to search for evidence of orbital modulation, an orbital phase was assigned to each $\gamma$-ray photon based on the timing ephemeris reported by \citet{gui12b}, using the \fermi\ plug-in for \textsc{Tempo2}\footnote{See http://fermi.gsfc.nasa.gov/ssc/data/analysis/user/Fermi\_plug\_doc.pdf}, taking an aperture radius of $1\degr$. No significant evidence of orbital modulation at a $>2\sigma$ level was found by employing the H-test \citep{djb10}, which is consistent with the results from \citet{gui12b}. To access the effect of the sky exposure as a function of time throughout the observation, a lightcurve was created with a bin size equal to the one-tenth of the orbital period of the binary system, the exposure was calculated with the tool {\it gtexposure}. The summed exposure in each bin was found to deviate from the mean value by less than $2\%$, suggesting that the variation in exposure has a minute effect on the orbital lightcurve. To avoid bias due to the alignment of the first bin with the orbital phase zero, the bins were shifted by phases of 0.2, 0.4, 0.6 and 0.8. The results after shifting the lightcurve do not change our conclusion.

\begin{figure}
\plotone{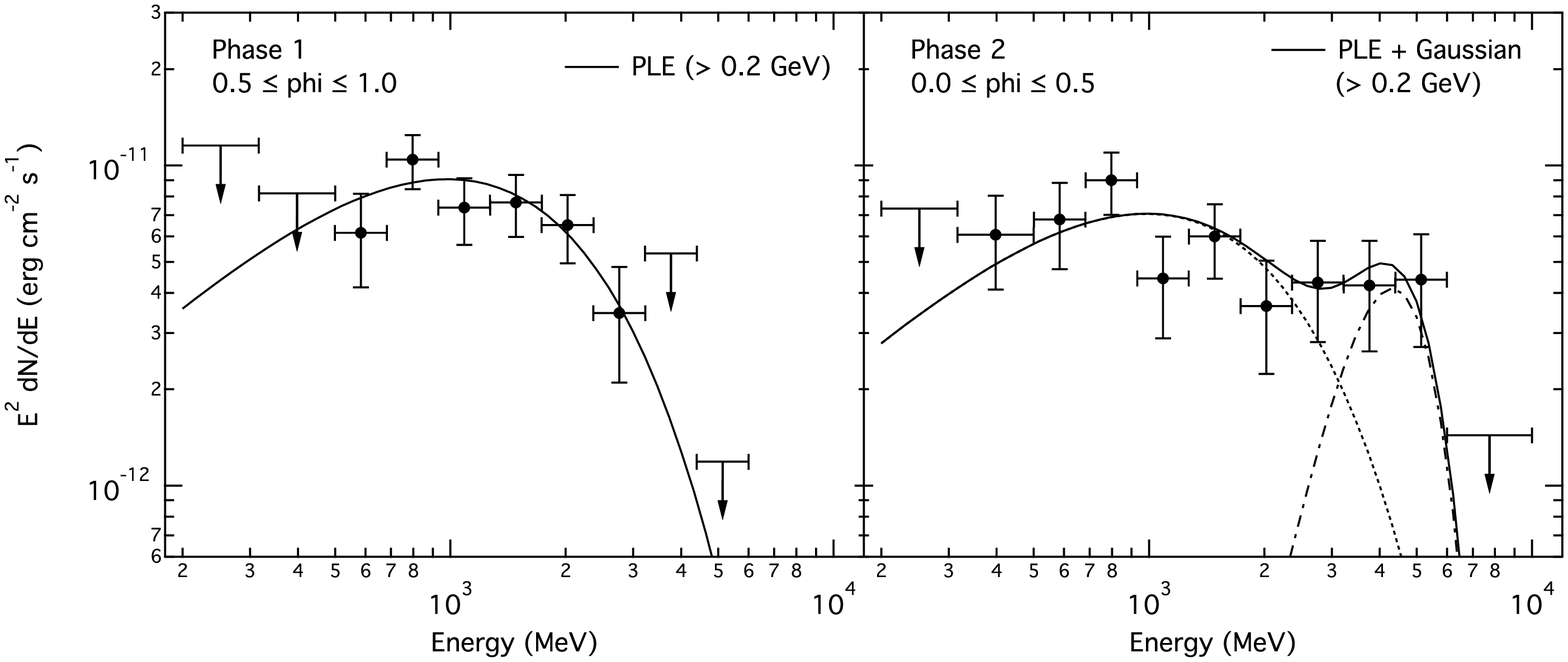}
\caption{Spectral energy distributions of $\gamma$-ray emission from PSR~B1957+20. Data points were derived from likelihood fitting of individual energy bins, in which a simple PL is used to model the data. 90\% upper limits were calculated for any energy bin in which the detection significance is lower than $3\sigma$. {\bf Left:} Spectrum averaged over Phase~1. The solid line shows the best-fit PLE model from fitting the data above $0.2$ GeV. {\bf Right:} Spectrum averaged over Phase~2. The solid line represents the fitted two-component model, with the PLE component shown as a dashed line and the Gaussian component shown as a dash-dotted line.}
\label{B1957sed}
\end{figure}

Recently, \citet{hua12} have discovered orbital-modulated X-ray emission from the PSR B1957+20 system. To investigate whether the $\gamma$-ray spectral properties also vary with the orbital phase ($\phi$), we divided an orbital cycle into two parts: half of the orbit centered at the superior conjunction (hereafter ``Phase~1'') and the other half containing the inferior conjunction (hereafter ``Phase~2''). Unbinned likelihood analysis was performed for both regions with the help of {\it gtlike}. A spectral-spatial model containing all other sources reported in the Second \fermi\ source catalog (\citealt{nol12}, 2FGL hereafter), within $10\degr$ from the center of the ROI was used for source modelling, resulting in a total of 13 point sources in the model. All the 2FGL sources were assumed to have the spectral type as suggested in the catalog. The spectral parameters of all sources within the ROI were set to be free, while the parameters for sources outside the region were kept fixed. We also included the Galactic diffuse model ({\tt gal\_2yearp7v6\_v0.fits}) as well as the isotropic spectral template ({\tt iso\_p7v6source.txt}) and allow their normalizations to be free.
Since the $\gamma$-ray spectra of pulsars in general can be characterized by a power-law with an exponential cut-off (PLE) model, $dN/dE = N_0 (E / 0.2~\rm{GeV})^{-\Gamma} \exp{[-(E / E_c)]}$ \citep{abd10}, where $N_0$ is a normalization factor, $\Gamma$ is the photon index and $E_c$ is the cutoff energy, we performed fitting of the spectrum average over each orbit segment with a PLE model. For Phase~1, the best-fit parameters are a photon index of $\Gamma_1 = 0.83 \pm 0.56$ and a cut-off energy of $E_{\rm cutoff,1} = 0.84 \pm 0.30$ GeV, while for Phase~2, $\Gamma_2 = 1.58 \pm 0.30$ and $E_{\rm cutoff,2} = 2.28 \pm 0.88$ GeV. A summary of the fitting results is presented in Table~\ref{fitresults}. The difference in cut-off energies in the two phases suggests that the spectrum of Phase~2 extends to higher energies than that of Phase~1. Moreover, the observed spectrum of Phase~1 is fitted by the PLE model with a Test-Statistic (TS) value of 205, while it is fitted by a PL model with a TS value of 168. This indicates that the PLE model is more favored over the PL model in describing the spectrum of Phase~1 at a $\sim 6.1\sigma$ confidence level. For Phase~2, the TS values of the best-fit PLE and PL models are 161 and 143 respectively, indicating that the PLE model is not significantly favored over the PL model. The spectral energy distributions (SEDs) for both orbital phases are presented in Figure~\ref{B1957sed}. We computed the extrapolated fluxes in the energy range 0.1 -- 300 GeV for both Phase~1 and Phase~2, which are also presented in Table~\ref{fitresults}. The values of the fluxes are consistent with that reported in \citet{gui12b} within uncertainties. We note that in the source model, the best-fit log-parabola model for the source 2FGL J1949.7+2405 yields a curvature coefficient $\beta \approx 0.2$ -- $0.3$, which is different from that given in the 2FGL catalog, where the value of $\beta$ is fixed at $1$. However, it is pointed out in \citet{nol12} that such highly curved spectrum is not necessarily robust, possibly due to the densely populated region near the Galactic plane. Moreover, a likelihood analysis using an ROI of radius 15 deg (see below) yields a similar result, suggesting that the deviation could be justifiable.

Since the radius of the point-spread function (PSF) is relatively large near the low energy bound of $200~{\rm MeV}$, the spectral results may be affected by the wings of the PSFs of surrounding sources. Thus, we performed binned likelihood analysis with an ROI of radius 15\degr around PSR~B1957+20. The spectral fitting results are presented in Table~\ref{fitresults}. These results are consistent with those obtained using an ROI of radius 5\degr. We also note that the difference in the best-fit cut-off energies in Phase~1 and Phase~2 agrees with the conclusion on the difference in the spectral properties of the two orbital phases.

\begin{figure}
\epsscale{0.6}
\plotone{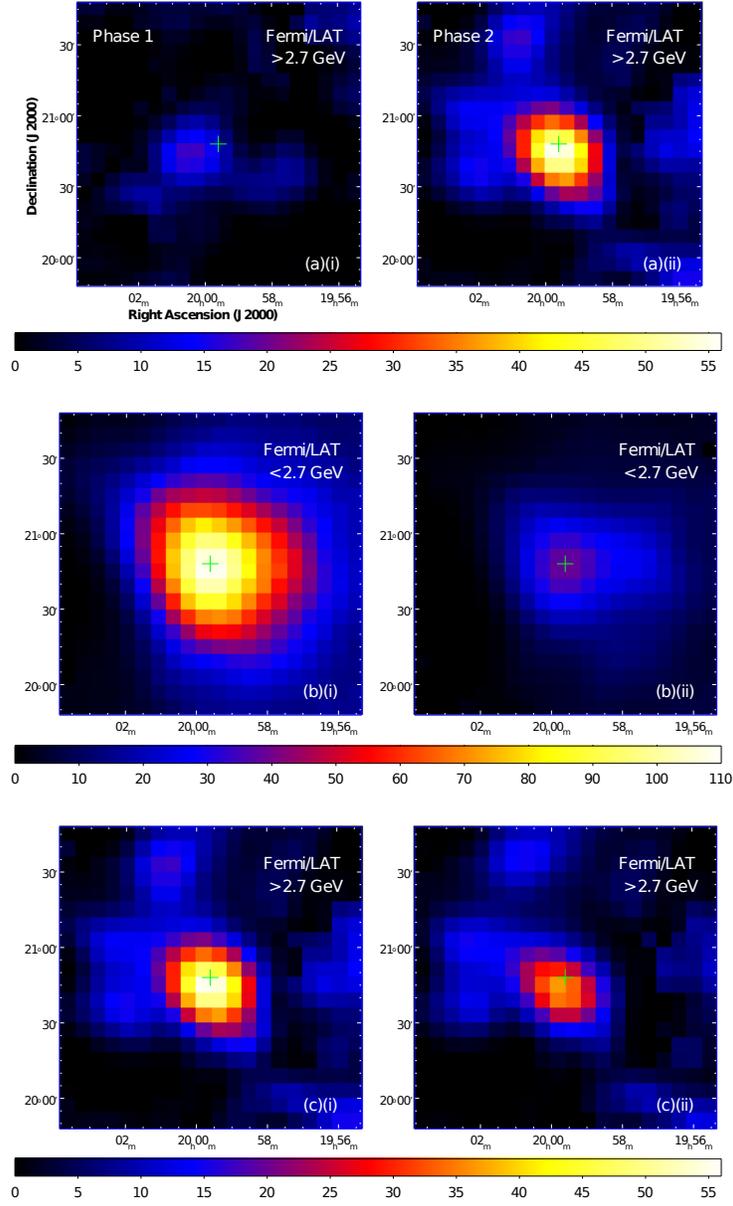}
\caption{Test-statistic (TS) maps of $2\degr \times 2\degr$ regions centered at the position of PSR B1957+20 (labeled by green crosses). The color scale below each pair of images is used to indicate the TS values. {\bf (a):} (i) TS map at energy $> 2.7$ GeV using only photons in Phase~1. (ii) Same as (a)(i) but using only photons in Phase~2. {\bf (b):} (i) TS map at $< 2.7$ GeV for Phase~2. (ii) Same as (b)(i) but with data within full width at half maximum of the pulsation peaks removed (see text). {\bf (c):} Same as (b) but with energy $> 2.7$ GeV.}
\label{tsmaps}
\end{figure}

By comparing the $\gamma$-ray spectra, we speculate that the spectrum for Phase~2 could be described by two components, one being magnetospheric, which has also been well-established by \citet{gui12b}, and the other component above $\sim 3$~GeV, coming from the interaction between the pulsar wind and the companion star. To describe this component while avoiding bias towards any emission scenario, we adopted a simple Gaussian profile $dN/dE = A \exp{-[(E - \bar{E})^2 / \sigma_{G}^2]}$ to fit the data along with a PLE model with the spectral index and the cutoff energy fixed at the best-fit values derived from Phase~1. The best-fit two-component model is presented as the solid line in the right panel of Figure~\ref{B1957sed}. The best-fit parameters for the Gaussian component are $\bar{E} = 3.76 \pm 0.59$~GeV and $\sigma_{G} = 1.10 \pm 0.39$~GeV. To access the significance of emission of this extra component, we performed likelihood analysis using the data at energies above $\bar{E} - \sigma_{G} \approx 2.7$ GeV by assuming a simple PL model, alleviating the effect of the spectral model on the significance. For Phase~2, a TS value of $55$ was achieved, corresponding to a significance of $\sim7\sigma$, whereas for Phase~1, a TS value of $14$ was obtained, indicating that the detection significance is below $4\sigma$ at $> 2.7$ GeV for Phase~1. We computed the TS maps using {\it gttsmap} at energies $> 2.7$ GeV for both Phase~1 and Phase~2, which are shown in Figure~\ref{tsmaps}(a). The difference in the TS values for the two orbital phases is readily observable by comparing the maps.

\begin{deluxetable}{ccccccccc}
\tabletypesize{\scriptsize}
\tablewidth{0pc}
\tablecaption{Summary of spectral fitting results of the two orbital phases}
\startdata
\hline\hline
Orbital & ROI radius (\degr) & Spectral & $\Gamma$ & $E_c$ & $\bar{E}$ & $\sigma_G$ & Flux\tablenotemark{a} \\
phase & (analysis type) & model & & (GeV) & (GeV) & (GeV) & ($10^{-9}~\rm{ph~cm^{-2}~s^{-1}}$) \\
\hline
0.5 -- 1.0 & 5 (unbinned) & PLE & $0.83\pm0.56$ & $0.84\pm0.30$ & \nodata & \nodata & $8.95\pm1.79$ \\
(Phase 1) & & & & & & & ($12.8\pm4.42$) \\
 & 15 (binned) & PLE & $1.21\pm0.57$ & $1.04\pm0.46$ & \nodata & \nodata & $8.76\pm1.99$ \\
 & & & & & & & ($14.3\pm5.94$) \\
\hline
0.0 -- 0.5 & 5 (unbinned) & PLE & $1.58\pm0.30$ & $2.28\pm0.88$ & \nodata & \nodata & $8.47\pm1.57$ \\
 (Phase 2) & & & & & & & ($15.0\pm4.52$) \\
 & 5 (unbinned) & PLE + Gaussian & 0.83\tablenotemark{b} & 0.84\tablenotemark{b} & $3.76\pm0.59$ & $1.10\pm0.39$ & $7.19\pm0.85$ \\
 & & & & & & & ($10.2\pm1.22$) \\
 & 15 (binned) & PLE & $1.88\pm0.19$ & $3.17\pm1.15$ & \nodata & \nodata & $8.12\pm1.28$ \\
 & & & & & & & ($16.6\pm3.77$) \\
 & 15 (binned) & PLE + Gaussian & 1.21\tablenotemark{b} & 1.04\tablenotemark{b} & $3.80\pm0.70$ & $1.00\pm0.54$ & $6.76\pm0.96$ \\
 & & & & & & & ($10.9\pm$1.57) \\
\enddata
\tablenotetext{a}{\footnotesize Integrated photon flux in the energy range 0.2 -- 300 GeV. The values in brackets represent the 0.1 -- 300 GeV extrapolated photon flux.}
\tablenotetext{b}{\footnotesize Model parameters without quoted errors are fixed at the values given.}
\label{fitresults}
\end{deluxetable}

\begin{figure}
\epsscale{0.8}
\plotone{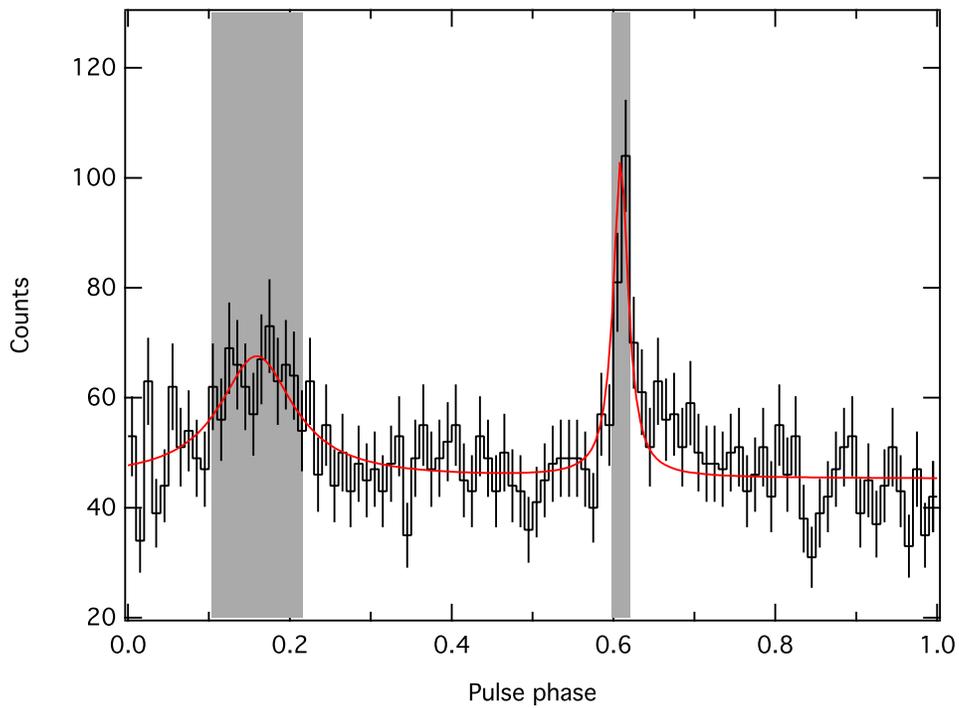}
\caption{Pulse profile of the $\gamma$-ray emission from PSR B1957+20, produced by using data above $0.2$ GeV and within $1^{\circ}$ from the position of the pulsar. The red solid line represents the fitted Lorentzian functions. The shaded regions indicate the pulse phases within the full widths at half maximum of the two peaks.}
\label{pulselc}
\end{figure}

In addition, we investigate the relation between the rotation phase of the pulsar and the variation of spectral characteristics in the two orbital phases by removing the pulsed component. A pulse profile of PSR B1957+20 was constructed by phase-folding the photons with energies above $0.2$ GeV and within an angular distance of $1^{\circ}$ from the position of the pulsar. The gamma-ray peaks in the profile was then fitted with two Lorentzian functions. Data were excluded within the full widths at half-maximum of the fitted peaks, which consist of phases 0.103 -- 0.215 and 0.598 -- 0.620. The pulse profile and the fitted Lorentzian functions are illustrated in Figure~\ref{pulselc}. We compare the significance of emission in Phase~2 below and above $2.7$~GeV, before and after removing the pulsation peaks. At energies below $2.7$~GeV, the TS value decreased from 105 to 38, while at energies above $2.7$~GeV, the TS value decreased from 55 to 36. Figure~\ref{tsmaps}(b) and (c) show the TS maps for a visual comparison before and after the removal of the pulsed component.

The above results indicate the presence of emission above 2.7 GeV, which is significantly detected in Phase~2 but not in Phase~1. Hence, the $\gamma$-ray emission from PSR B1957+20 is dependent on the orbital phase of the binary system. Moreover, the removal of the pulsed emission component has caused a greater decrease in the detection significance at energies below 2.7 GeV than above, implying the majority of the emission of 2.7 GeV in Phase~2 are not produced inside the pulsar magnetosphere, but in the intra-binary region.

\begin{figure}
\plotone{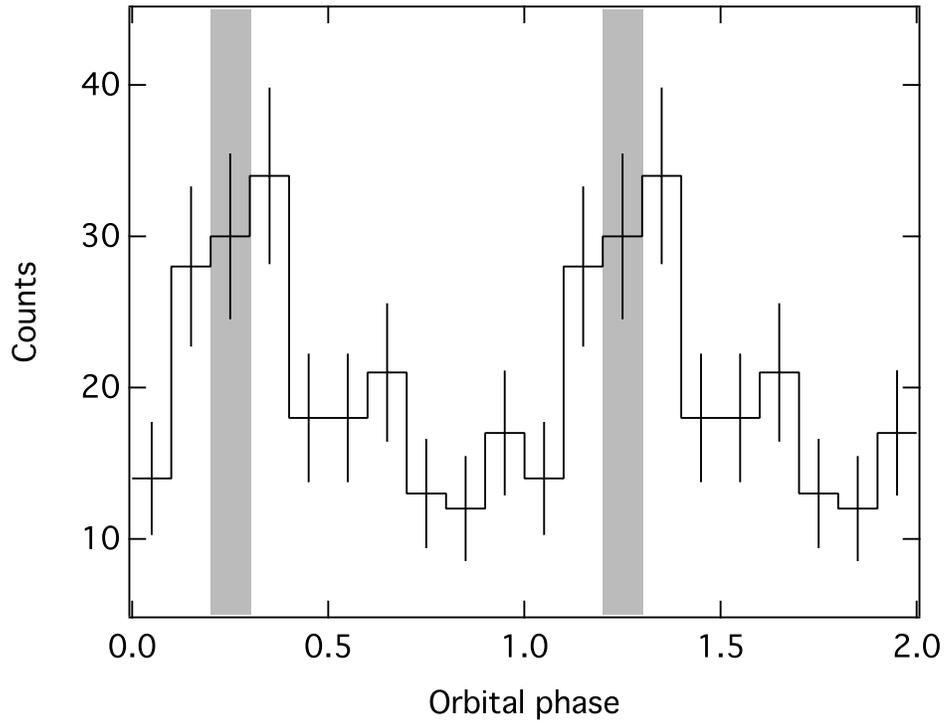}
\epsscale{0.8}
\caption{$\gamma$-ray lightcurve of PSR~B1957+20 folded at the orbital period using the $\fermi$ plug-in for \textsc{Tempo2}, with the optimized size of aperture of $0\fdg965$. Two orbits are shown for clarity. The shaded regions correspond to the phase of radio eclipse, which is the center of Phase~2.}
\label{foldedlc}
\end{figure}

Since the best-fit two-component model suggests that photons at $\gtrsim 2.7$~GeV may be modulated with the orbital phase, we constructed a phase-folded lightcurve and derived the significance of the modulation. To allow for enough photon statistics we selected events with energies greater than $2.7$ GeV, the size $r$ of the aperture. The best profile ($r = 0\fdg965$) is presented in Figure~\ref{foldedlc}, with an H-test Test Statistic of 19 \citep{djb10}, corresponding to a significance of $\sim 3.3\sigma$. The post-trial significance, associated with 21 trials on the aperture radius, is $\sim 2.3\sigma$.

\section{Discussion}
We have reported the evidence of orbital-modulated GeV $\gamma$-ray emissions from the black widow system PSR B1957+20. We have fitted the spectrum of Phase~2 (half of the orbit centered at the inferior conjunction) using best-fit PLE model of Phase~1 plus the Gaussian component with $\bar{E} = 3.76 \pm 0.59$ GeV and $\sigma_{G} = 1.10 \pm 0.39$ GeV. The additional component in the spectrum has been found with $\sim 7\sigma$ confidence level. We have also shown that the orbital modulation above $\sim 3$~GeV can be found with a $\sim 2.3\sigma$ confidence level.  

It can be expected that the PLE component below 3~GeV will be dominated by the emissions from the pulsar magnetosphere. For the additional component above 3~GeV in Phase~2, we will speculate the possibility of the inverse-Compton process of a ``cold'' ultra-relativistic pulsar wind. In this section, we will calculate the expected spectra of the emissions from pulsar magnetosphere and from the pulsar wind. 

\subsection{Outer gap emissions}
For the magnetospheric emission, we apply the two-layer outer gap model explored by \citet{wtc10,wtc11}. The dashed lines in Figure~\ref{modelspe} represent the calculated phase-averaged spectra of the curvature radiation from the outer gap. The results are for the inclination angle of $\alpha=60^{\circ}$ and the viewing angle of $\xi=80^{\circ}$, respectively, which were assumed to explain the pulse profile and the spectrum. A larger viewing angle is required to explain the phase separation ($\sim 0.45$ orbital phase) of the two peaks  in the light curve of the pulsed emissions. The assumed viewing geometry is consistent with that used in \citet{gui12b}. For the present outer gap model, the $\gamma$-ray luminosity is expected to be order of $L_{\rm gap}\sim f^{3}L_{sd}$, where $f$ is fractional gap thickness, which is ratio of the gap thickness to the light cylinder radius at the light cylinder, and $L_{sd}\sim 1.6\times 10^{35}~{\rm erg~s^{-1}}$ \citep{man05}\footnote{http://www.atnf.csiro.au/research/pulsar/psrcat} is the spin down power of PSR~B1957+20. We note that because the period time derivatives before and after correction of Shklovskii effect \citep{shk70} due to proper motion are $\dot{P}\sim 1.68\times 10^{-20}$ and $\sim 1.13\times 10^{-20}$ \citep{man05}, respectively, the spin down power and hence our main conclusion are not changed much by the Shklovskii effect. For the fractional gap thickness, we apply the outer gap model controlled by the magnetic pair-creation process near the stellar surface \citep{tak10}, which implies $f\sim 2.5(P/0.1 {\rm s})^{1/2}\sim 0.3$ for PSR B1957+20. The right panel in Figure~\ref{modelspe} compares the model spectrum of the outer gap and the observed $\gamma$-ray  spectrum in off-peak orbital phase in Figure~\ref{foldedlc}, where we expect the magnetospheric emission dominates. We can see in Figure~\ref{modelspe} that the emissions from the outer gap can explain well the $Fermi$ data in off-peak orbital phase.

\subsection{Pulsar wind emissions}
It has been thought that a ``cold'' pulsar wind, which will be mainly composed of the electrons and positron pairs, is injected into the interstellar space by the pulsars. The pulsar wind is accelerated beyond the light cylinder to a Lorentz factor $\Gamma_W\sim 10^{5-7}$  \citep{ken84}. If the cold ultra-relativistic pulsar wind interacts with the inter-stellar medium, the kinetic energy of the pulsar wind is converted into the internal energy at the shock. The shocked pulsar wind can emit X-rays via synchrotron radiation and is observed as the pulsar wind nebula. The un-shocked cold pulsar wind is dark in the X-ray bands, because the wind does not emit synchrotron photons. However, it can been thought that the cold ultra-relativistic pulsar wind emits very high-energy $\gamma$-rays by the inverse-Compton process between the cold pulsar wind and the background soft-photons from the pulsar magnetosphere or from the companion in the binary system. In fact, studies have been carried out on the $\gamma$-ray emissions from the cold pulsar wind of the Crab pulsar \citep{abk12} and of the $\gamma$-ray binary PSR B1259-63 \citep{kha07,tak09,kha12}.

For PSR B1957+20 black widow system, the temperature of the heating side of companion is as high as $T_{\rm eff}\sim 8300$~K \citep{rey07}. Roughly speaking the luminosity of the inverse-Compton process may be estimated by $L_{\rm IC}\sim (1-\cos\theta_{0})\sigma_{T}n_{ph}a_{s}L_{sd}\sim 5\times 10^{-3}(1-\cos\theta_0)(T_{\rm eff,3}/8.3)^3 R_{s,10}^{2}a_{s,11}^{-1}L_{\rm sd}$, where $\theta_{0}$ is the collision angle between the pulsar wind and the soft photon, $R_s\sim 10^{10}R_{s,10}$~cm is the Roche lobe radius and $a_{s}=10^{11}a_{s,11} \,{\rm cm}$ is the separation between the pulsar and the companion. In addition, $n_{ph}\sim \sigma_{sb}T_{\rm eff}^3R_s^2/k_{b}ca_s^2$ with $k_B$ being the Boltzmann constant, is the typical number density of the soft photon field, and $T_{\rm eff,3}=(T_{\rm eff}/1000~\rm K)$. Because the typical luminosity of outer gap emissions is of the order of $L_{\rm gap}\sim f^3L_{\rm sd}\sim 0.03L_{\rm sd}$, the inverse-Compton process of pulsar wind may produce $\gamma$-ray radiation with a flux level that is of the same order of magnitude as that of the gap emissions, if the inverse-Compton process occurs with head-on collision, which can happen around the inferior conjunction, because the companion is located between the pulsar and the Earth. In fact, we have measured  the significant excess of the $\gamma$-ray emissions above 3 GeV during  the inferior conjunction passage. The typical energy of the scattered photons is $\sim 3\Gamma_W^2kT_{\rm eff}\sim 5~{\rm GeV}(T_{\rm eff,3}/8.3)(\Gamma_W/5\times 10^{4})^2$. 

We perform a more detail calculation for the inverse-Compton process between the cold ultra-relativistic pulsar wind and the soft-photon field from the companion. The emissivity of the inverse-Compton process of a particle may be expressed by \citep{beg87,tak12}, 
\begin{equation}
\frac{dP_{\rm IC}}{d\Omega}={\mathcal D}^2\int_0^{\theta_c}(1-\beta\cos\theta_0) I_b/h\frac{d\sigma'_{\rm KN}}{d\Omega'}d\Omega_0,
\end{equation}
where $d\sigma'_{\rm KN}/d\Omega'$ is the differential Klein-Nishina cross section, $\beta=\sqrt{\Gamma_W^2-1}/\Gamma_W$, $\mathcal{D}=\Gamma_W^{-1}(1-\beta\cos\theta_1)^{-1}$ with $\theta_1$ being the angle between the direction of the particle motion and the propagating direction of the scattered photon, $h$ is the Planck constant, $I_b$ is the background photon field and $\theta_c$ expresses the angular size of star as seen from the emission point.

We assume that all pulsar wind's particles have a Lorentz factor of $\Gamma_{W}=4\times 10^4$. We consider that the energy flux of the pulsar wind depends on the colatitude ($\theta$) measured from the rotation axis as $L_{W}(\theta)\propto \sin^2\theta$ \citep{bog02,lyu02}, and that total power of the wind corresponds to the spin down power. We also assume that half of the energy flux is carried by the kinetic energy of the pulsar wind. The fitting of the optical modulation using ELC model implies that the inclination angle of the orbital plane with respect to the sky is $i\sim 65^{\circ}$ \citep{rey07}. In the present calculation, we assume $i=67^{\circ}$, and the viewing angle $\xi=80^{\circ}$ respective to the rotation axis of PSR B1957+20. The distance to the source has been estimated as $d\sim 2.5\pm 1$~kpc \citep{fst88,gui12b}. In the present calculation, we apply $d=2$~kpc. We note that the observed flux can be reproduced more easily with a larger inclination angle ($i$) and a smaller distance than the estimated values within the tolerance of their uncertainties.

\begin{figure}
\epsscale{.80}
\plotone{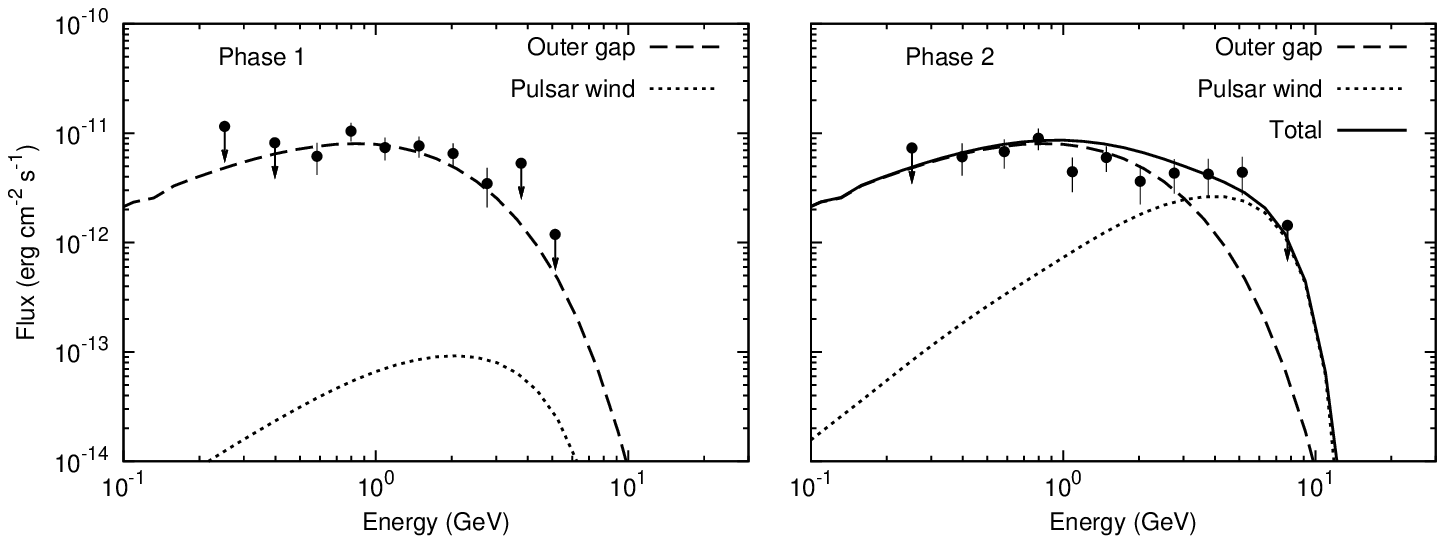}
\caption{The spectrum averaged over Phase~1 (left panel) and Phase~2 (right panel). The dashed and the dotted lines are the spectra of the emissions from the outer gap and the cold ultra-relativistic pulsar wind, respectively. The solid line represents the total contribution from the components.}
\label{modelspe}
\end{figure}

\begin{figure}
\epsscale{.80}
\plotone{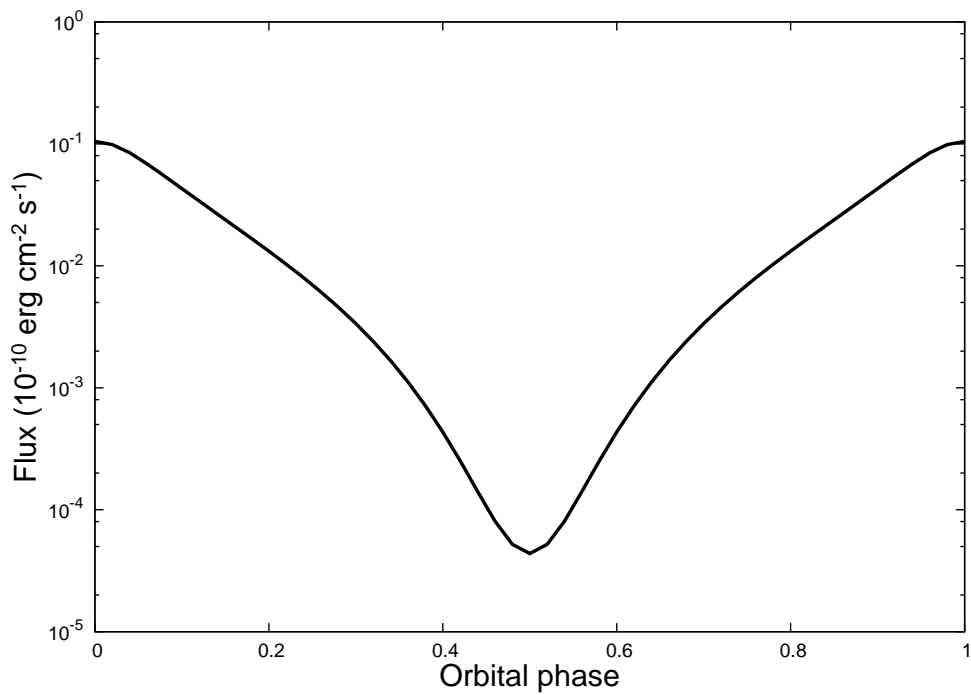}
\caption{Calculated orbital modulation for $\gamma$-ray emissions from the inverse-Compton process of a cold ultra-relativistic pulsar wind. The phase 0 corresponds to the inferior conjunction.}
\label{light}
\end{figure}

The spectra of the inverse-Compton process of the pulsar wind are described by the dotted lines in Figure~\ref{modelspe}, and the temporal behavior is summarized in Figure~\ref{light}, where the phase zero corresponds to the inferior conjunction. We have expected that the pulsar passes the inferior conjunction at the center of X-ray peak phase (Phase~2). At the inferior conjunction passage, because the companion is located between the pulsar and the Earth, the inverse-Compton process occurs with head-on collision rather than tail-on collision. We found that the emissions from inverse-Compton process dominates the curvature radiation from the outer gap in the spectrum  above 3~GeV, and can explain the observed excess in the \fermi\ data at the peak orbital phase, as the left panel in Figure~\ref{modelspe} indicates. In the off-peak phase (Phase~1), because the inverse-Compton process occurs with tail-on collision rather than head-on, the inverse-Compton process is suppressed. Hence, only magnetospheric emissions contribute to the observed emissions, as shown in right panel of Figure~\ref{modelspe}.

As we have seen, the inverse-Compton process of the cold ultra-relativistic pulsar wind can explain the observed extra component above 3~GeV in the orbital phases around the inferior conjunction. Hence, we would like to emphasize that the PSR~B1957+20 binary system is (1) a plausible candidate manifesting the emissions from the cold ultra-relativistic wind, and (2) a candidate of the new class of $\gamma$-ray binary, in the sense that the $\gamma$-ray binary is composed of a MSP and a low mass star, and showing both magnetospheric and pulsar wind emissions. 

It is unlikely that the extra component above 3~GeV during the inferior conjunction passage is originated from the inverse-Compton process of {\it shocked} pulsar wind. We may estimate the magnetic field of the pulsar wind after the shock by $B=3 L_{\rm sd}^{1/2} a_s c^{-1/2} \sigma_{W}^{1/2}(1+\sigma_W)^{-1/2}$ \citep{tak09}, where $\sigma_W$ is the ratio of the magnetic energy to the particle energy at the shock. The energy density of the thermal radiation from the companion is at most $U_{ph}=\sigma_{\rm SB}T_{\rm eff}^4/c$. Hence, the ratio of powers of the synchrotron radiation and the inverse-Compton process of the shocked pulsar wind is estimated as $U_B/U_{ph}\sim 10L^{1/2}_{\rm sd,35}~(T_{\rm eff,3}/8.3)^{-4}\sigma_W(1+\sigma_W)$, where $L_{\rm sd,35}=(L_{\rm sd,35}/10^{35}~{\rm erg~s^{-1}})$. On the other hand,  the ratio of the observed fluxes of the orbital modulated X-ray emissions ($\sim 10^{-13}{\rm erg~cm^{-2}~s^{-1}}$, Huang et al. 2012) and GeV emissions ($\sim 10^{-11}~{\rm erg~cm^{-2}~s^{-1}}$) are $\sim 10^{-2}$. Hence, the inverse-Compton process of the shocked pulsar wind will not explain the  emissions above $\sim 3$~GeV observed at the inferior conjunction passage.

Finally, we remark that the black widow pulsars discovered in \emph{Fermi} unidentified sources are good targets for searching orbital phase-dependent $\gamma$-ray spectra. In particular, the black widow candidates 2FGL~J2339.6-0532 and 2FGL~J1311.7-3429 show orbital modulation in optical, and it is evident that the former candidate shows X-ray modulation as well. Hence, they are an excellent probe for $\gamma$-ray emission with both magnetospheric and pulsar wind components.

In summary, despite the limitation from the significance of each individual approach, the pieces of evidence as a whole has allowed us to conclude the detection of orbital modulated $\gamma$-rays from the PSR~B1957+20 system in \fermi-LAT data. We have shown that the significant emissions above 3~GeV appears around inferior conjunction, while the emissions below 3~GeV are steady and are described by the pulsar emissions. We expect that the modulated emissions above 3~GeV is originated from the inverse-Compton scattering of the thermal radiation of the companion star off the cold ultra-relativistic pulsar wind.

\acknowledgements 
We thank P.~Ray for useful discussions and suggestions for \fermi\ data analysis. We express our appreciation to the anonymous referee for useful comments and suggestions. This project is supported by the National Science Council of the Republic of China (Taiwan) through grant NSC100-2628-M-007-002-MY3 and NSC100-2923-M-007-001-MY3. CYH is supported by the National Research Foundation of Korea through grant 2011-0023383. A.~K.~H.~K.~gratefully acknowledges support from a Kenda Foundation Golden Jade Fellowship. E.~M.~H.~W., J.~T.~and K.~S.~C.~are supported by a GRF grant of HK Government under HKU700911P.

\end{document}